\documentclass[aps,prl,preprint,showpacs]{revtex4} 
\usepackage{graphicx}
\usepackage{amssymb}
\usepackage{epsfig}

\begin{document}

\title{Echo Phenomena in Large Systems of Coupled Oscillators}

\author{Edward Ott, John H. Platig, Thomas M. Antonsen and Michelle Girvan}

\affiliation{University of Maryland, College Park, MD 20742}


\begin{abstract}
As exemplified by the Kuramoto model, large systems of coupled
oscillators may undergo a transition to phase coherence with
increasing coupling strength. It is shown that below the critical
coupling strength for this transition such systems may be expected
to exhibit `echo' phenomena: a stimulation by two successive
pulses separated by a time interval $\tau $ leads to the
spontaneous formation of response pulses at a time $\tau$, $2\tau
$, $3\tau \ldots$, after the second stimulus pulse.  Analysis of
this phenomenon, as well as illustrative numerical experiments,
are presented.  The theoretical significance and potential uses of
echoes in such systems are discussed.
\end{abstract}

\maketitle

{\bf Large systems consisting of many coupled oscillators for
which the individual natural oscillator frequencies are different
naturally occur in a wide variety of interesting applications. As
shown by Kuramoto, such systems can undergo a type of dynamical
phase transition such that as the coupling strength is raised past
a critical value, global synchronous collective behavior results.
In this paper we show that another interesting, potentially
useful, behavior of these systems also occurs {\it below} the
critical coupling strength. Namely, we demonstrate that these
systems exhibit {\it echo} phenomena:  If a stimulus pulse is
applied at time $t=0$, followed by a second stimulus pulse at time
$t=\tau $, then pulse echo responses can appear at $t=2\tau ,3\tau
,\ldots$.  This phenomenon depends on both nonlinearity and memory
inherent in the oscillator system, the latter being a consequence
of the continuous spectrum of the linearized system.}

\section{I.  Introduction}

Due to their occurrence in a wide variety of circumstances,
systems consisting of a large number of coupled oscillators with
different natural oscillation frequencies have been the subject of
much scientific interest\cite{pikovsky01,strogatz04}. Examples
where the study of such systems is thought to be relevant are
synchronous flashing of fireflies\cite{buck88} and chirping of
crickets\cite{walker69}, synchronous cardiac pacemaker
cells\cite{michaels87}, brain function\cite{singer93},
coordination of oscillatory neurons governing circadian rhythms in
mammals\cite{yamaguchi02}, entrainment of coupled oscillatory
chemically reacting cells\cite{kiss05}, Josephson junction circuit
arrays\cite{wiesenfeld95}, etc.  The globally-coupled,
phase-oscillator model of Kuramoto\cite{kuramoto84,acebron05}
exemplifies the key generic feature of large systems of coupled
oscillators. In particular, Kuramoto considered the case where the
distribution function of oscillator frequencies was monotonically
decreasing away from its peak value, and he showed that, as the
coupling strength $K$ between the oscillators is increased through
a critical coupling strength $K_c$, there is a transition to
sustained global cooperative behavior. In this state $(K>K_c)$ a
suitable average over the oscillator population (this average is
often called the `order parameter') exhibits steady macroscopic
oscillatory behavior. For $K<K_c$ a stimulus may transiently
induce macroscopic oscillations, but the amplitude of these
coherent oscillations (i.e., the magnitude of the order parameter)
decays exponentially to zero with increasing time\cite{acebron05}.
In the present paper we consider the Kuramoto model in the
parameter range $K<K_c$, and we demonstrate that `echo' phenomena
occur for this system. The basic echo phenomenon can be described
as follows: A first stimulus is applied at time $t=0$, and the
response to it dies away; next, a second stimulus is applied at a
later time, $t=\tau $, and its response likewise dies away; then
at time $t=2\tau $ (also possibly at $n\tau $, for $n=3,4,\ldots$)
an echo response spontaneously builds up and then decays away. An
illustrative example is shown in Fig.~1, which was obtained by
\begin{figure}[h]
\includegraphics[scale=.8]{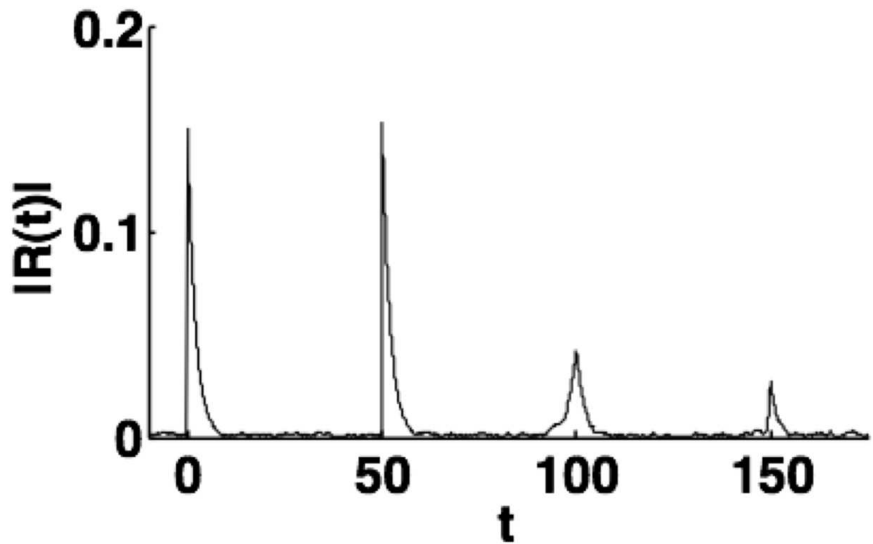}
\caption{Illustration of the echo phenomenon. Stimuli at times
$t=0$ and $t=\tau $ lead to direct system responses which rapidly
decay away followed by echo responses that can arise at times
$2\tau $, $3\tau , \ldots $. The `response' plotted on the
vertical axis is the magnitude of the complex valued order
parameter, Eq.~(8). See Sec.~IV for details of this computation.}
\end{figure}
numerical simulation (see Sec.~IV for details). In order for this
phenomenon to occur, the system must have two fundamental
attributes, nonlinearity and memory. Nonlinearity is necessary
because the response seen in Fig.~1 is not the same as the sum of
the responses to each of the individual stimulus pulses in the
absence of the other pulse (which is simply the decay that occurs
immediately after the individual stimuli, without the echo).
Memory is necessary in the sense that the system state after the
decay of the second pulse must somehow encode knowledge of the
previous history even though the global average of the system
state, as represented by the order parameter, is approximately the same as
before the two pulses were applied.

Echo phenomena of this type, occurring in systems of many
oscillators having a spread in their natural oscillation
frequencies, have been known for a long time. The first example
was the `spin echo' discovered in 1950 by Hahn\cite{hahn50}, where
the distribution of frequencies resulted from the position
dependence of the precession frequency of nuclear magnetic dipoles
in an inhomogeneous magnetic field. [The spin echo forms the basis
for modern magnetic resonance imaging (MRI).] Subsequently, echoes
for cyclotron orbits of charged particles in a magnetic field have
been studied for the cases in which the distribution in frequency
was due to magnetic field inhomogeneity\cite{gould65},
relativistic dependence of the particle mass on its
energy\cite{ott70}, and Doppler shifts of the cyclotron
frequency\cite{porkolab68}. Another notable case is that of plasma
waves, where the frequency distribution results from the Doppler
shift of the wave frequency felt by charged particles with
different streaming velocities\cite{oneil68}. Although echo
phenomena are well-known in the above settings, they have so far
not received attention in the context of the Kuramoto model and
its many related situations. It is our purpose in the present
paper to investigate that problem. Two possible motivations for
our study of echoes in the Kuramoto model are that they provide
increased basic understanding of the model and also that they may
be of potential use as a basis for future diagnostic measurements
of related systems (see Sec.~V).

In what follows, Sec.~II will give a formulation of the model
problem that will be analyzed in Sec.~III and numerically
simulated in Sec.~IV, while Sec.~V will provide a discussion of
the implications of the results obtained.

\section{II. Formulation}

We consider the basic Kuramoto model supplemented by the addition
of a $\delta $-correlated noise term $n(t)$ and two impulsive
stimuli, one at time $t=0$, and the other at time $t=\tau $,
\begin{equation}
d\theta _i/dt=\omega _i+K/N\sum ^N_{j=1}\sin (\theta _j-\theta
_i)-h(\theta _i)\Delta (t)+n(t) \ ,
\end{equation}
\begin{equation}
\Delta (t)=\hat d_0\delta (t)+\hat d_1\delta (t-\tau ) \ ,
\end{equation}
\begin{equation}
\langle n(t)n(t')\rangle =2\xi \delta (t-t') \ ,
\end{equation}
\begin{equation}
h(\theta )=\sum _nh_ne^{in\theta } \ , \ \ h_n=h^*_{-n} \ , \ \
h_0=0 \ ,
\end{equation}
where $h^*_{-n}$ denotes the complex conjugate of $h_{-n}$.  In
the above $\theta _i(t)$ represents the angular phase of
oscillator $i$, where $i=1,2,\ldots ,N\gg 1$; and $\omega _i$ is
the natural frequency of oscillator $i$ where we take $\omega _i$
for different oscillators (i.e., different $i$) to be distributed
according to some given, time-independent distribution function
$g(\omega )$, where $g(\omega )$ has an average frequency $\bar
\omega =\int \omega g(\omega )d\omega $, is symmetric about
$\omega =\bar \omega $, and monotonically decreases as $|\omega
-\bar \omega |$ increases.

To motivate the impulsive stimuli term, consider the example of a
population of many fireflies, and imagine that the stimuli at
$t=0$ and at $t=\tau $ are external flashes of light at those
times, where the constants $\hat d_0$ and $\hat d_1$ in Eq.~(2)
represent the intensity of these flashes. We hypothesize that a
firefly will be induced by a stimulus flash to move its flashing
phase toward synchronism with the stimulus flash. Thus a firefly
that has just recently flashed will reset its phase by retarding
it, while a firefly that was close to flashing will advance its
phase. The amount of advance or retardation is determined by the
`reset function', $h(\theta )$. Since the reset function $h(\theta
)$ depends on properties of the fireflies, we do not specify it
further. Let $\theta ^+_i$ and $\theta ^-_i$ represent the phases
of oscillator $i$ just after and just before a stimulus flash at
$t=0$ or $t=\tau$. Then we have from Eq.~(1) that
\begin{equation}
\int ^{\theta ^{+}_{i}}_{\theta _i^-} \frac{d\theta }{h(\theta
)}=\hat d_p ; \ \ p=0,1 \ .
\end{equation}
Letting $F(\theta )=\int ^\theta d\theta /h(\theta )$, we obtain
\begin{equation}
\theta ^+_i=F^{-1}(\hat d_p+F(\theta ^-_i)) \ .
\end{equation}
In our subsequent analysis in Sec.~III, we will for convenience
assume that $\hat d_p$ is small, in which case $(\theta
^+_i-\theta ^-_i)$ is small, and we can use the approximation,
\begin{equation}
\theta _i^+\cong \theta ^-_i+\hat d_ph(\theta ^-_i); \ p=0,1 \ .
\end{equation}

Following Kuramoto we introduce the complex valued order parameter
$R(t)$,
\begin{equation}
R(t)=\frac{1}{N} \sum ^N_{j=1} e^{i\theta _j(t)} \ ,
\end{equation}
in terms of which Eq.~(1) can be rewritten as
\begin{equation}
d\theta _i/dt=\omega _i+(K/N)Im[e^{-i\theta _i}R(t)]-h(\theta
_i)\Delta (t)+n(t) \ .
\end{equation}

In our analysis in Sec.~III we will take the limit $N\rightarrow
\infty$ useful for approximating the situation where $N\gg 1$. In
that limit it is appropriate to describe the system state by a
continuous distribution function $f(\theta ,\omega ,t)$, where
\begin{equation}
\int ^{2\pi }_0f(\theta ,\omega ,t)\frac{d\theta}{2\pi }=1 \ ,
\end{equation}
and the fraction of oscillators with angles and natural
frequencies in the ranges $(\theta , \theta +d\theta )$ and
$(\omega ,\omega +d\omega )$ is $f(\theta ,\omega ,t)g(\omega
)d\omega d\theta /2\pi $. The conservation of the number of
oscillators then gives the time evolution equation for $f(\theta
,\omega ,t)$,
\begin{equation}
\frac{\partial f}{\partial t}+\frac{\partial}{\partial \theta
}\left\{ f\left[ \omega +KIm(R(t)e^{-i\theta })-h(\theta )\Delta
(t)\right]\right\}=\xi \frac{\partial ^2f}{\partial \theta ^2} \ ,
\end{equation}
\begin{equation}
R^*(t)=\int d\omega f_1(\omega ,t)g(\omega ) \ ,
\end{equation}
where $R^*$ denotes the complex conjugate of $R$, $f(\omega
,\theta ,t)\equiv 1$ for $t<0$, and, in writing Eq.~(12), $f_1$
represents the $e^{i\theta }$ component of the Fourier expansion
of $f(\omega ,\theta ,t)$ in $\theta $,
\begin{equation}
f(\omega ,\theta ,t)=\sum ^{+\infty}_{n=-\infty }f_n(\omega
,t)e^{in\theta } \ ,
\end{equation}
with $f_0=1$, $f_n=f^*_{-n}$. As seen in Eq.~(11), the effect of
the noise term in Eq.~(1) is to introduce diffusion in the phase
angle $\theta $ whose strength is characterized by the phase
diffusion coefficient $\xi $.

In Sec.~III we will solve Eqs.~(11) and (12) for the case $d_p\ll
1$, thus demonstrating the echo phenomenon as described in Sec.~I.
In Sec.~IV we will present numerical solutions of Eq.~(1) for
large $N$.

\section{III.  Analysis}
\subsection{A. Amplitude expansion}
In order to proceed analytically we use a small amplitude
expansion and obtain results to second order (i.e., up to
quadratic in the small amplitude). This will be sufficient to
obtain the echo phenomenon. We introduce a formal expansion
parameter $\epsilon $, as follows,
\begin{equation}
f=1+\epsilon f^{(1)}+\epsilon ^2f^{(2)}+\mathcal{O}(\epsilon ^3) \
;
\end{equation}
$\hat d_p=\epsilon d_p $ for $p=0$, $1$; $R=\epsilon
R^{(1)}+\epsilon ^2R^{(2)}+\mathcal{O}(\epsilon ^3)$;
$R^{(m)*}=\int gf_1^{(m)}d\omega $; where $f^{(m)}=\Sigma
_nf_n^{(m)}\exp (in\theta )$. (Although we formally take $\epsilon
\ll 1$, when we finally get our answers, the results will apply
for $\epsilon =1$ and $d_p=\hat d_p$, if $\hat d_p\ll 1$.)
\subsection{B. Order $\epsilon $}
In linear order (i.e., $\mathcal{O}(\epsilon))$, by multiplying
Eq.~(11) by $\exp (-i\theta )d\theta $ and integrating over
$\theta $, we have for the component of $f^{(1)}$ varying as
$e^{i\theta }$,
\begin{equation}
\frac{\partial f_1^{(1)}}{\partial t}+(i\omega +\xi
)f_1^{(1)}=\frac{K}{2} R^{(1)*}+ih_1\Delta (t) \ , \ \
R^{(1)*}(t)=\int f^{(1)}gd\omega \ ,
\end{equation}
where $f_1^{(1)}(\omega ,t)=0$ for $t<0$ and $R^{(1)*}$ is the
complex conjugate of $R^{(1)}$. Due to the delta function term on
the right hand side of Eq.~(15), $ih_1d_0\delta (t)$, at the
instant just after the first delta function (denoted $t=0^+$),
$f_1^{(1)}$ jumps from zero just before the delta function
(denoted $t=0^-$) to the value $f_1^{(1)}(\omega ,0^+)=ih_1d_0$.
Making use of this observation, in Appendix I we solve Eq.~(15)
for $0<t<\tau $, with the result that, for $K<K_c$,
\begin{equation}
f_1^{(1)}(\omega ,t)=A(\omega )e^{-(i\omega +\xi )t}+\ \ ({\rm a\
more\ rapidly\  exponentially\  decaying\  component)} \ ,
\end{equation}
where
\begin{equation}
A(\omega )=ih_1d_0/D[-(i\omega +\xi )] \ ,
\end{equation}
\begin{equation}
D(s)=1-\frac{K}{2}\int ^{+\infty}_{-\infty} \frac{g(\omega
)d\omega }{s+\xi +i\omega } \ , \ {\rm for} \ Re(s)>0 \ ,
\end{equation}
and $D(s)$ for $Re(s)\leq 0$ is defined from Eq.(18) by analytic
continuation. Since Eq.~(16) applies for $0<t<\tau $, we have that
just before the application of the second delta function stimulus
$(t=\tau ^-)$,
\begin{equation}
f_1^{(1)}(\omega ,\tau ^-)\cong A(\omega )e^{-(i\omega +\xi )\tau
} \ ,
\end{equation}
where we have neglected the second term on the right hand side of
Eq.~(16) on the basis that, due to its more rapid exponential
decay, it is small compared to the first term.

Solutions of $D(s)=0$ govern the stability of the state with
$R^{(1)}=0$. Let $s=s_0$ denote the solutions of $D(s)=0$ with the
largest real part. If $Re(s_0)<0$ the state $R^{(1)}=0$ is stable,
and a perturbation away from $R^{(1)}=0$ decays to zero with
increasing $t$ at the exponential rate $Re(s_0)$. If $Re(s_0)>0$,
then the perturbation grows and $R^{(1)}$ eventually saturates
into a sustained nonlinear state of coherent cooperative
oscillatory behavior\cite{kuramoto84,acebron05}. In general,
$Re(s_0)$ is an increasing function of the coupling constant $K$,
and $Re(s_0)\stackrel{>}{<}0$ for $K\stackrel{>}{<}K_c$, where
$K_c$ is a critical value that depends on $\xi$ and $g(\omega )$.
Throughout this paper we shall be considering only the case
$K<K_c$ for which $Re(s_0)<0$.

It is instructive to consider $\xi =0$. In that case, the first
term in Eq.~(16) is of constant magnitude in time, but, as time
$t$ increases, it oscillates more and more rapidly as a function
of $\omega $. Because of this increasingly rapid variation in
$\omega $, the contribution of this term to $R^{(1)*}(t)=\int
gf_1^{(1)}d\omega $ decays in time (see Appendix I), and it does
so at the same time-asymptotic rate as the contribution from the
second more rapidly exponentially decaying contribution in Eq.~(16). Thus the order parameter
magnitude decays away, but the distribution function $f_1^{(1)}$
can still have a component (the first term in Eq.~(16)) due to the
pulse that has not decayed away. A similar conclusion applies for
$\xi >0$ provided that $\xi $ is substantially less than the
damping for the second term in Eq.~(16). This is the source of the
`memory' referred to in Sec.~I. It is also worth noting that the
first term in Eq.~(16) can be thought of as the manifestation of
the continuous spectrum of the Kuramoto problem, discussed in
detail in Ref.\  \cite{strogatz92}. Thus the echo phenomenon that
we derive subsequently can be regarded as an observable
macroscopic consequence of the continuous spectrum, where by
`macroscopic' we mean that the effect can be seen through
monitoring of the order parameter without the necessity of other
more detailed knowledge of the distribution function.

It is also of interest to consider $f_n^{(1)}$ for $n\geq 2$. From
Eq.~(11) we obtain for $|n|\geq 2$
\begin{equation}
\frac{\partial f_n^{(1)}}{\partial t}+(in\omega +n^2\xi
)f_n^{(1)}=inh_n\Delta (t) \ ,
\end{equation}
which does not have any contribution from the order parameter,
$R$. For $\tau >t>0$, Eq.~(20) yields
\begin{equation}
f_n^{(1)}(\omega ,t)=inh_nd_0\exp [-(in\omega +n^2\xi )t]\ ,
\end{equation}
for $0 < t < \tau $, which, similar to the first term on the right
hand side of Eq.~(16), also oscillates increasingly more rapidly
with $\omega $ as $t$ increases. At time $t=\tau ^-$ Eq.~(21)
yields
\begin{equation}
f_n^{(1)}(\omega ,\tau ^-)=inh_nd_0\exp [-(in\omega +n^2\xi )\tau
] \ ,
\end{equation}
for $|n|\geq 2$.
\subsection{C. Order $\epsilon ^2$}
Now proceeding to $\mathcal{O}(\epsilon ^2)$ and again (as done in
obtaining Eq.~(15)) taking the $e^{i\theta }$ component of
Eq.~(11), we have
\begin{equation}
\frac{\partial f_1^{(2)}}{\partial t}+(i\omega +\xi
)f_1^{(2)}-\frac{1}{2}KR^{(2)*}=-i\left\{
\frac{K}{2i}f_2^{(1)}R^{(1)}-\Delta (t)\sum
^{+\infty}_{n=-\infty}h_{-(n-1)}f^{(1)}_n\right\}
\end{equation}
where $R^{(1,2)*}(t)=\int ^{+\infty}_{-\infty}g(\omega
)f_1^{(1,2)}(\omega ,t)d\omega $. The above equation is linear in
$f_1^{(2)}$ and is driven by several inhomogeneous terms appearing
on the right hand side of Eq.~(23) that are quadratic in first
order quantities. Since we are interested in the components of
$f_1^{(2)}$ that result in echoes, and since, by our previous
discussion, we expect that the echoes depend on the presence of
{\it both} stimulus delta functions (i.e., the delta function
$\delta (t)$ of strength $d_0$ and the delta function $\delta
(t-\tau )$ of strength $d_1$), we are interested in the component
of $f_1^{(2)}$ that is proportional to the product $d_0d_1$ for
$t>\tau $. We denote this component $f^{(2)}_{1,e}$, where the
subscript $e$ stands for `echo'. From Eq.~(23) we see that for
$t>\tau $, the $f^{(2)}_{1,e}$ component of $f_1^{(2)}$ satisfied
the following initial value problem
\begin{equation}
\frac{\partial f_{1,e}^{(2)}}{\partial t}+(i\omega +\xi
)f^{(2)}_{1,e}-\frac{1}{2}KR_e^{(2)*}=0 \ ,
\end{equation}
\begin{equation}
f^{(2)}_{1,e}(\omega ,\tau ^+)=id_1\sum
^{+\infty}_{n=-\infty}h_{-(n-1)}f^{(1)}_n(\omega ,\tau ^-) \ ,
\end{equation}
\begin{equation}
R_e^{(2)*}(t)=\int ^{+\infty}_{-\infty} g(\omega
)f^{(2)}_{1,e}(\omega ,t)d\omega \ .
\end{equation}
Since $f^{(1)}_n(\omega ,\tau^- )$ is proportional to $d_0$ (see
Eqs.~(19) and (22)), we see that the solution of Eqs.~(24)--(26)
for $f_{1,e}^{(2)}$ and $R_e^{(2)}$ will indeed be proportional to
$d_0d_1$ as desired.

We solve Eqs.~(24)--(26) by taking Laplace transforms,
\begin{equation}
\hat f^{(2)}_{1,e}(\omega ,s)=\int ^\infty _\tau
e^{-st}f_{1,e}^{(2)}(\omega ,t)dt \ ,
\end{equation}
\begin{equation}
\hat R^{(2)}_{e*}(s)\equiv \int ^\infty _\tau
e^{-st}R_e^{(2)*}(t)dt \ ,
\end{equation}
in terms of which we obtain from Eq.~(24)
\begin{equation}
\hat f^{(2)}_{1,e}(\omega ,s)=\hat R^{(2)}_{e*}\frac{K/2}{s+\xi
+i\omega }+\frac{f^{(2)}_{1,e}(\omega ,\tau ^+)e^{-s\tau }}{s+\xi
+i\omega } \ .
\end{equation}
Multiplying Eq.~(29) by $g(\omega )d\omega $ and integrating from
$\omega =-\infty$ to $\omega =+\infty$, then yields
\begin{equation}
\hat R^{(2)}_{e*}(s)=\frac{e^{-s\tau }}{D(s)}\int
^{+\infty}_{-\infty}\frac{f^{(2)}_{1,e}(\omega ,\tau ^+)}{s+\xi
+i\omega }g(\omega )d\omega \ .
\end{equation}
To find $R_e^{(2)*}(t)$ we take the inverse Laplace transform,
\begin{equation}
R_e^{(2)*}(t)=\frac{1}{2\pi i}\int ^{+i\infty +\eta
}_{-i\infty+\eta }e^{st}\hat R_{e*}^{(2)}(s)ds \ , \eta >0 \ .
\end{equation}
For the purposes of evaluating the integral (31), we recall that
$D(s)=0$ has roots whose real parts correspond to the exponential
decay rate of a response to an initial stimulus toward the $R=0$
state. Thus, as before in our discussion of the linear response
(see Eq.~(16)), any poles at the roots of $D(s)=0$ give
contributions that we assume decay substantially faster with
increasing $t>\tau $ than the diffusion induced exponential decay
rate $\xi $. Since we are interested in echoes that we will find
occur for $t=2\tau ,3\tau ,\ldots $, we neglect contributions to
Eq.~(31) from such poles. Thus it suffices to consider only the
contribution to Eq.~(31) from the pole at $s+\xi +i\omega =0$.
Hence Eqs.~(30) and (31) yield
\begin{equation}
R_e^{(2)*}(t)\cong \int ^{+\infty}_{-\infty}e^{-(i\omega +\xi
)(t-\tau )}\frac{f^{(2)}_{1,e}(\omega ,\tau ^+)}{D[-(i\omega +\xi
)]} g(\omega )d\omega \ .
\end{equation}
\subsection{D. Echoes}
In order to see how Eq.~(32) results in echoes, we recall our
previous results, Eqs.~(25), (19) and (21) for
$f_{1,e}^{(2)}(\omega ,\tau ^+)$, and combine them to obtain
\begin{equation}
f^{(2)}_{1,e}(\omega ,\tau ^+)=d_0d_1h_2h_1^*\frac{\exp (i\omega
\tau -\xi \tau )}{D^*[-(i\omega +\xi )]}-d_0d_1\sum _{|n|\geq
2}nh_nh^*_{n-1}\exp [-(in\omega +n^2\xi )\tau ] \ ,
\end{equation}
where we have used $h_0=0$, \ $h_n=h^*_n$,
$f^{(1)}_{-1}=f_1^{(1)*}$, and the first term on the right side of
Eq.~(33) corresponds to $n=-1$ in Eq.~(25). Putting Eq.~(33) into
Eq.~(32), we see that we have an integral of a sum over terms with
exponential time variations of the form
\begin{equation}
\exp\{-i\omega [t-(1-n)\tau ]\}\exp \{ -\xi [t+(n^2-1)\tau \} \ .
\end{equation}
Considering the first exponential in Eq.~(34), we see that, for
large values of $|t-(1-n)\tau |$, there is rapid oscillation of
the integrand with $\omega $, and the integral can therefore be
expected to be near zero. However, such rapid oscillation is
absent near the times $t=(1-n)\tau $, at which a large value of
$R_e^{(2)*}$ will occur. Since $t>\tau $, the relevant times occur
for $n<-1$; e.g., for $n=-1$, we get an echo at $t=2\tau$; for
$n=-2$, we get an echo at $t=3\tau $; etc. Therefore, we
henceforth replace the summation over $|n|\geq 2$ in Eq.~(33) by a
summation from $n=-\infty$ to $n=-2$.
\subsection{E. Evaluation for Lorentzian frequency distribution
functions}

We now consider the case of a Lorentzian frequency distribution,
\begin{equation}
g(\omega )=g_L(\omega )\equiv \frac{1}{\pi }\frac{\Delta}{(\omega
-\bar \omega )^2+\Delta ^2}=\frac{1}{2\pi i}\left\{
\frac{1}{\omega -(\bar \omega +i\Delta )}-\frac{1}{\omega -(\bar
\omega -i\Delta )}\right\} \ .
\end{equation}
The right-most expression for $g_L(\omega )$ makes clear that,
when the previously real variable $\omega $ is analytically
continued into the complex plane, the function $g_L(\omega )$
results from the sum of two pole contributions, one at $\omega
=\bar \omega +i\Delta $, and one at $\omega =\bar \omega -i\Delta
$. The quantity $\bar \omega $ represents the average frequency of
the distribution, while $\Delta $ represents the width of the
distribution. Consideration of the Lorentzian will be particularly
useful to us because the integral (32) can be explicitly
evaluated, and also because our numerical experiments in Sec.~IV
will be for the case of a Lorentzian frequency distribution
function.

As a first illustration we consider the $n=-1$ term which results
in an echo at $t=2\tau $. We first evaluate $D(s)$ by inserting
the pole-form for $g_L(\omega )$ into Eq.~(18) and closing the
integration path with a large semicircle of radius approaching
infinity. This yields a single residue contribution to $D(s)$,
\begin{equation}
D(s)=1-\frac{K}{2}[s+\xi +i(\bar \omega -i\Delta )]^{-1} \ .
\end{equation}
Note that the solution of $D(s)=0$ occurs at
\begin{equation}
s=-i\omega -\left(\xi +\Delta -\frac{K}{2}\right) \ .
\end{equation}
According to our previous assumptions, we require $K<K_c\equiv
2(\Delta +\xi )$ so that the $R=0$ state is stable, and $(\Delta
-K/2)\tau - \xi \tau \gg 1$ so that we can neglect contributions
from the pole at the root $D(s)=0$ in our approximation of (31) by
(32). Using Eq.~(36) and the $n=-1$ contribution to
$f^{(2)}_{1,e}$ (i.e., the first term in (33)) in Eq.~(32) we
obtain for the echo term at $t=2\tau $ (denoted $R^{(2)*}_{2\tau
}(\epsilon )$),
\begin{equation}
R^{(2)*}_{2\tau }(t)=2ih^*_1h_2d_0d_1\Delta
\int^{+\infty}_{-\infty} \frac{d\omega }{2\pi i}\cdot
\frac{\exp[-i\omega (t-2\tau )-\xi t]}{[(\omega -\bar \omega
)-i(\Delta -\frac{K}{2})][(\omega -\bar \omega )+i( \Delta
-\frac{K}{2})]}\ .
\end{equation}
For $t>2\tau $ $(t<2\tau )$ the integrand exponentially approaches
zero as $Im(\omega )\rightarrow -\infty $ $(Im(\omega )\rightarrow
+\infty)$, and we can therefore close the integration path with a
large semicircle in the lower half $\omega $-plane (upper half
$\omega $-plane). Thus the integral (38) is evaluated from the
pole enclosed by the resulting path [i.e., the pole $\omega
=\omega _0-i(\Delta -\frac{K}{2})$ for $t>2\tau $, and the pole
$\omega =\omega _0+i(\Delta -\frac{K}{2})$ for $t<2\tau $],
\begin{equation}
R^{(2)*}_{2\tau}(t)=\frac{h^*_1h_2d_0d_1\Delta }{\Delta
-(K/2)}e^{-i\bar \omega (t-2\tau )-\xi t}e^{-(\Delta
-\frac{K}{2})|t-2\tau |} \ .
\end{equation}
From Eq.~(39) we see that we obtain an echo that is approximately
symmetric in shape about $t=2\tau $ (i.e., the envelope $\exp
[-(\Delta -K/2)|t-2\tau |]$) for $\xi \ll (\Delta -\frac{1}{2}K)$.

We can similarly evaluate the contribution $R^{(2)*}_{m\tau }(t)$
of echoes at $t=m\tau $ for $m=3,4,\ldots$. For example, the
result for the echo at $t=3\tau $ is
\begin{equation}
R^{(2)*}_{3\tau }=\frac{2h^*_2h_3d_0d_1\Delta }{\Delta
-(K/4)}e^{-\xi (3\tau +t)}e^{-i\bar \omega (t-3\tau )}E(t-3\tau )
\ ,
\end{equation}
\begin{equation}
E(t-3\tau )= \left\{ \begin{array}{ll} \exp [\Delta (t-3\tau )] \
, & {\rm for} \  t<3\tau \ , \\
\exp -[(\Delta -\frac{1}{2}K)(t-3\tau )] \ , & {\rm for} \
t>3\tau \ .
\end{array}
\right.
\end{equation}
Thus, in the case $\xi =0$, the shape of the pulse envelope
$E(t-3\tau )$ is asymmetric about $t=3\tau $, increasing at a more
rapid exponential rate (namely, $\Delta )$ as $t$ increases toward
$3\tau $, than the slower exponential rate of decrease (namely,
$\Delta -(K/2)$) as $t$ increases away from $3\tau $. This is in
contrast to the symmetrically shaped envelope $\exp [-(\Delta
-\frac{1}{2}K)|t-2\tau |]$ for the echo at $t=2\tau $.

In Appendix II we present an evaluation of $R^{(2)*}_{2\tau }(t)$
for the case of a Gaussian frequency distribution function,
\[
g(\omega )=g_G(\omega )\equiv [2\pi \Delta ^2]^{-1/2}\exp
[-(\omega -\bar \omega )^2/(2\Delta ^2)] \ .
\]
\subsection{F. The small coupling limit}
We now consider a general frequency distribution function
$g(\omega )$ but for the case where the coupling between
oscillators is small. That is, $K\ll \Delta $, where $\Delta $
denotes the frequency width of $g(\omega )$ about its mean value
$\omega =\bar \omega $. In this case a good approximation is
provided by setting $K=0$. Thus $D[-(i\omega +\xi )]\cong 1$ and
Eq.~(33) yields
\begin{equation}
f^{(2)}_{1,e}(\omega ,\tau ^+)=d_0d_1\sum ^\infty
_{n=1}nh^*_nh_{n+1}\exp [-(-in\omega +n^2\xi )\tau ] \ ,
\end{equation}
where we have replaced $n$ by $-n$ and used $h_n=h^*_{-n}$.
Inserting Eq.~(42) into Eq.~(32) we obtain
\begin{equation}
R_e^{(2)*}(t)=\sum ^\infty _{n=2}(n-1)d_0d_1h^*_{n-1}h_n\tilde
g(t-n\tau )e^{-[(n^{2}-1)\tau +t]\xi } \ ,
\end{equation}
where $\tilde g(t)$ is defined by
\begin{equation}
\tilde g(t)=\int ^{+\infty}_{-\infty}d\omega e^{-i\omega
t}g(\omega ) \ .
\end{equation}
Thus, for $K\ll \Delta $, the shape of the echoes at $t=2\tau
,3\tau ,\ldots$ is directly given by the Fourier transform (44) of
the frequency distribution function $g(\omega )$. Another point is
that with $K\rightarrow 0$, Eq.~(1) shows that the oscillators do
not interact, and the nonlinearity needed to produce the echo
phenomenon comes entirely from the stimulus function $h(\theta )$.

\section{IV. Simulations}
We have performed direct numerical simulations of the system (1)
with a Lorentzian oscillator distribution (see Eq.~(35)), $\bar
\omega =0$, $\Delta =1$ (corresponding to $K_c=2$), $\hat d_0=\hat
d_1$, $K=1$, $\tau =50$, and $\xi =0$. At $t=0^{-}$ we initialize
each phase $\theta _i$ for $i=1,2,\ldots ,N$ randomly and
independently with a uniform distribution in the interval $(0,2\pi
)$. We then apply the mapping given by Eq.~(7) with $\hat d_p=\hat
d_0$ to each $\theta _i$ in order to simulate the effect of the
delta function at $t=0$. Next we integrate Eq.~(1) for each
$i=1,2,\ldots ,N$ forward in time to $t=\tau ^{-}$, again apply
the mapping Eq.~(7) (but now with $\hat d_p=\hat d_1$), and we
then continue the integration. At each time step we also calculate
$R(t)$ using Eq.~(8). Figure 2 shows results for $\hat d_0=\hat
d_1=1/4$, and
\begin{figure}[h]
\includegraphics[scale=.75]{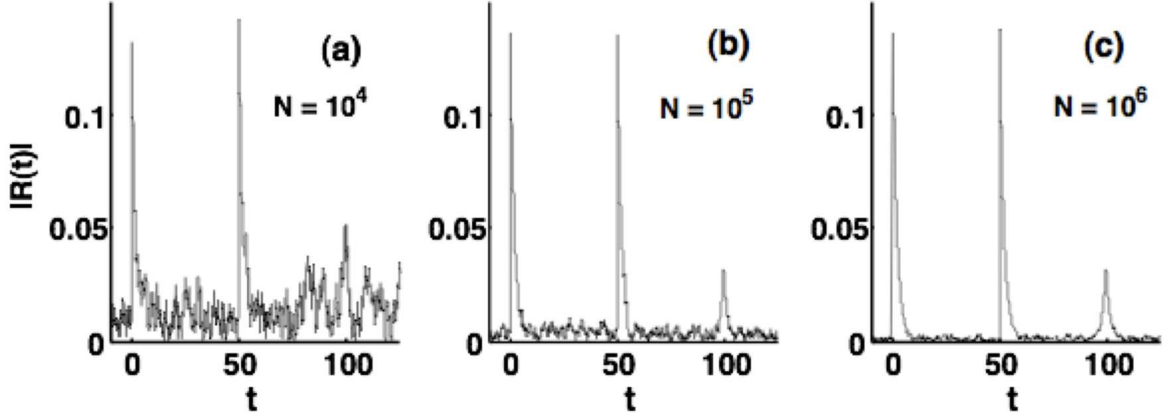}
\caption{$|R(t)|$ versus $t$ for (a) $N=10^6$, (b) $N=10^5$, (c)
$N=10^4$, and (d) $N=10^3$, showing the echo at $t\cong 2\tau $
and the increase of fluctuations at lower $N$.}
\end{figure}

\begin{figure}[h]
\includegraphics[scale=.75]{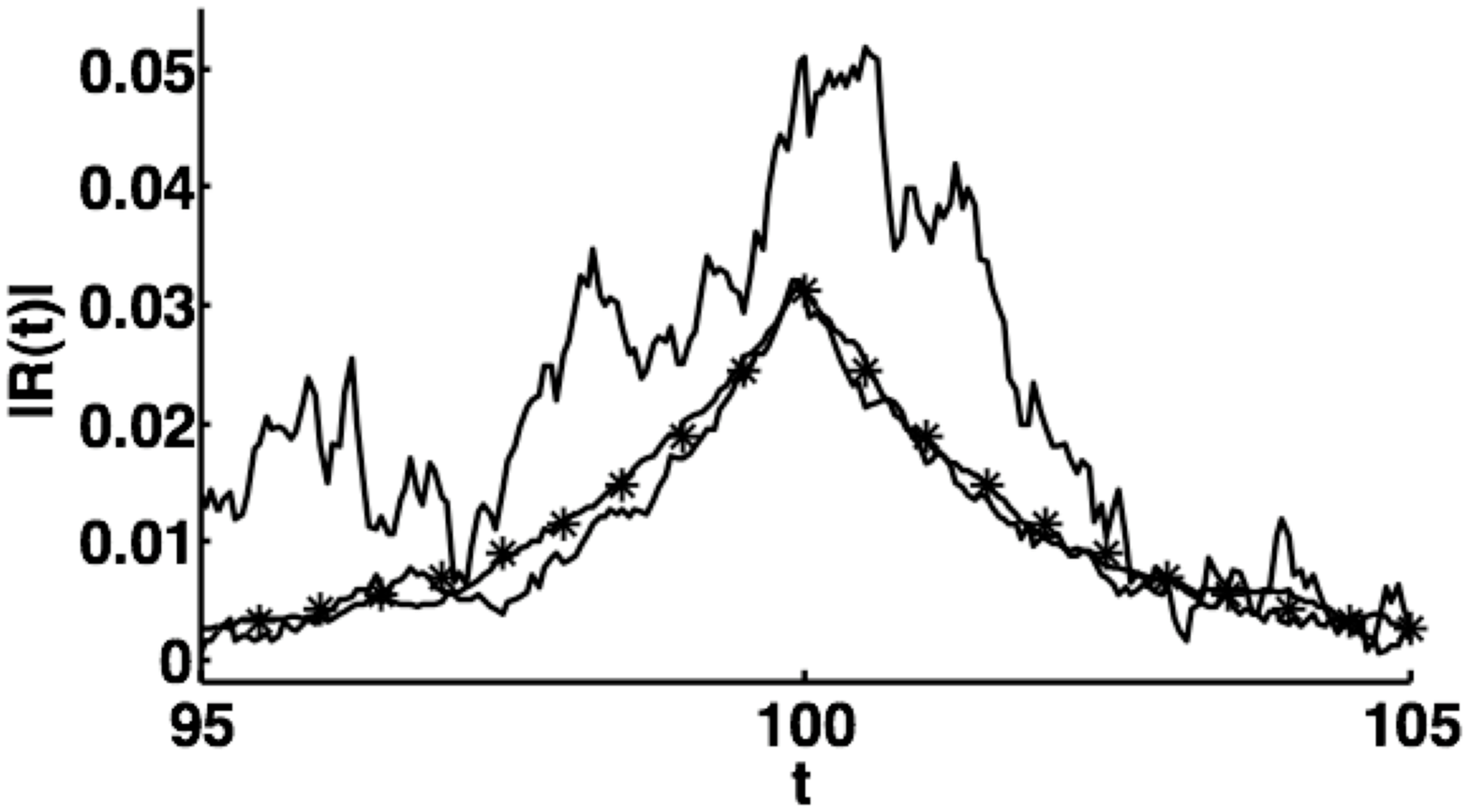}
\caption{$|R(t)|$ versus $t$ blown up around $t\cong 2\tau =200$,
for $N=10^6$, $10^5$, $10^4$, $10^3$ (solid curves) showing the
increase of fluctuations at lower $N$. The dotted curve is the
theoretical result from Eq.~(39) with $\xi =0$.}
\end{figure}
\[
h(\theta )=\sin \theta +\sin 2\theta \ ,
\]
for several different system sizes, $N=10^6$, $10^5$, and $10^4$.
 Figure 2(a--c) shows $|R(t)|$
versus $t$ for $0\leq t\leq 125$. The responses to the delta
functions at $t=0$ and $\tau $, as well as the echo at time
$t=2\tau $ are clearly illustrated. The effect of lower $N$ is to
increase the fluctuations making the echo somewhat less distinct.
We do not see any echo at $t=3\tau $. This is in agreement with
Eq.~(40), since $h_3=0$ for the $h(\theta )$ employed in these
computations. Figure 3 shows a blow-up of the numerically computed
echo around the time $t=2\tau $ for $N=10^6$, $10^5$, and $10^4$.
Also, plotted in Fig.~3 as asterisks is the result from our
theoretical calculation Eq.~(39). Reasonable agreement between the
theoretical and computed echo shapes is obtained, although the
agreement is somewhat obscured by fluctuation effects at the
smaller system sizes $(N)$. While our choice $\hat d_0=\hat
d_1=1/4$ might be regarded as questionable for applicability of
the small amplitude approximation $(\hat d_p\ll 1$, for $p=0,1$)
employed by Eq.~(7) and by our theory of Sec.~III, we have
nonetheless evidently obtained good agreement between the theory
and numerical experiment. Figure 4 illustrates the effect of
varying the driving amplitude for a network of size $N=10^4$. For
$\hat d_0=\hat d_1=1/8$ (Fig.~4(a)) the echo is swamped by the
noise and is not seen. For $\hat d_0=\hat d_1=1/4$ (Fig.~4(b),
same as 2(a)) the echo seems to have appeared, but because of the
noise, this conclusion is somewhat questionable. Finally, at the
larger driving of $\hat d_0=\hat d_1=1/2$, the echo is clearly
present.

Figures 5(a) and 5(b) show the effect of changing $h(\theta )$. In
particular, Fig.~5(a) shows numerical results for $\hat d_0=\hat
d_1=1/4$, $N=10^5$, and $h(\theta )=\sin \theta $, with all other
parameters the same as before. Since $h_2$ is now zero, Eq.~(39)
now predicts that there is no echo, in agreement with Fig.~5(a).
Figure 5(b) shows numerical results for $\hat d_0=\hat d_1=1/4$,
$N=10^5$, and
\[ h(\theta )=\sin \theta +\sin 2\theta +\sin 3\theta,\]
with all other parameters the same as before. Since $h_1$,
$h_2$ and $h_3$ are all nonzero, Eqs.~(39) and (40) now predict
echoes at both $t\cong 2\tau $ and at $t\cong 3\tau $, and this is
confirmed by Fig.~5(b).
\begin{figure}[h]
\includegraphics[scale=.5]{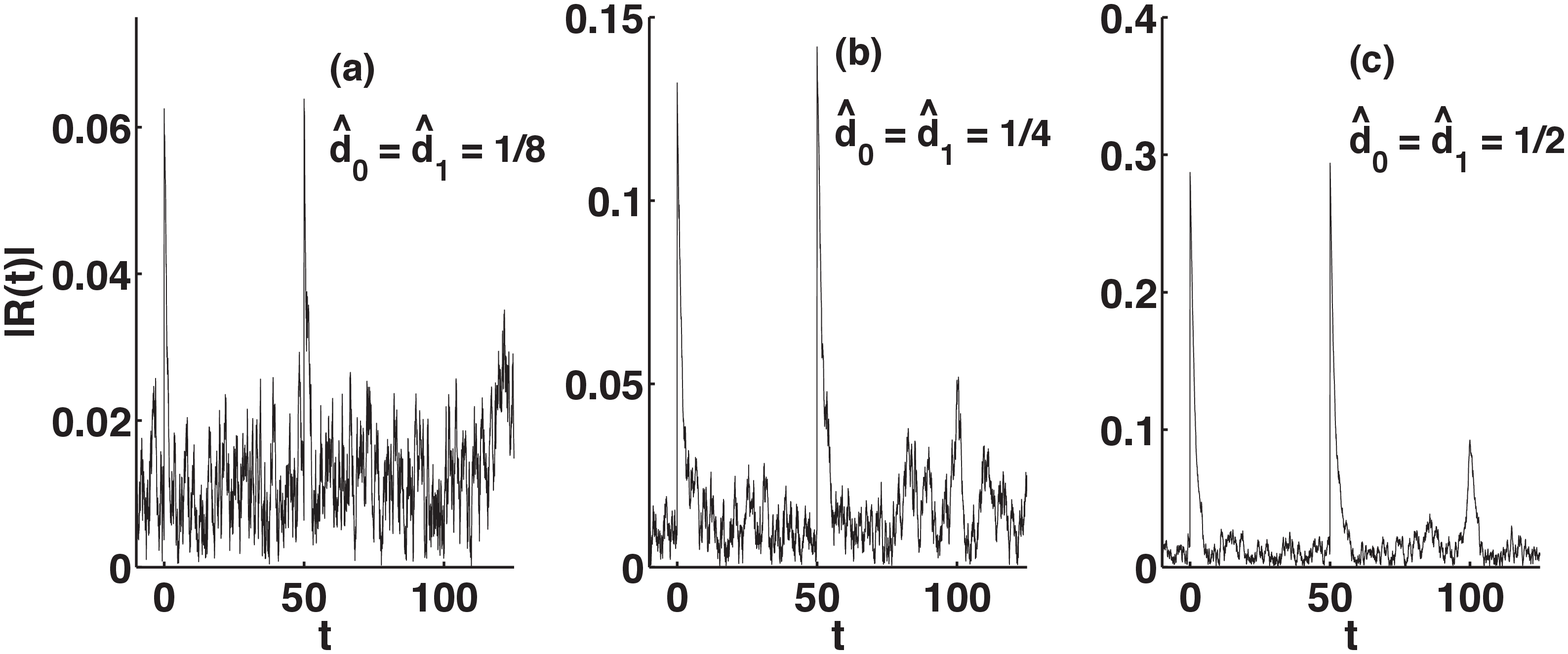}
\caption{Simulation of $10^5$ oscillators for $\tau =100$, $\hat
d_0=\hat d_1=1/3$, $K=1=\frac{1}{2}K_c$, $h(\theta )=\sin \theta
$. In this case, no echo at $t=2\tau =200$ is observed.}
\end{figure}

\begin{figure}[h]
\includegraphics[scale=.75]{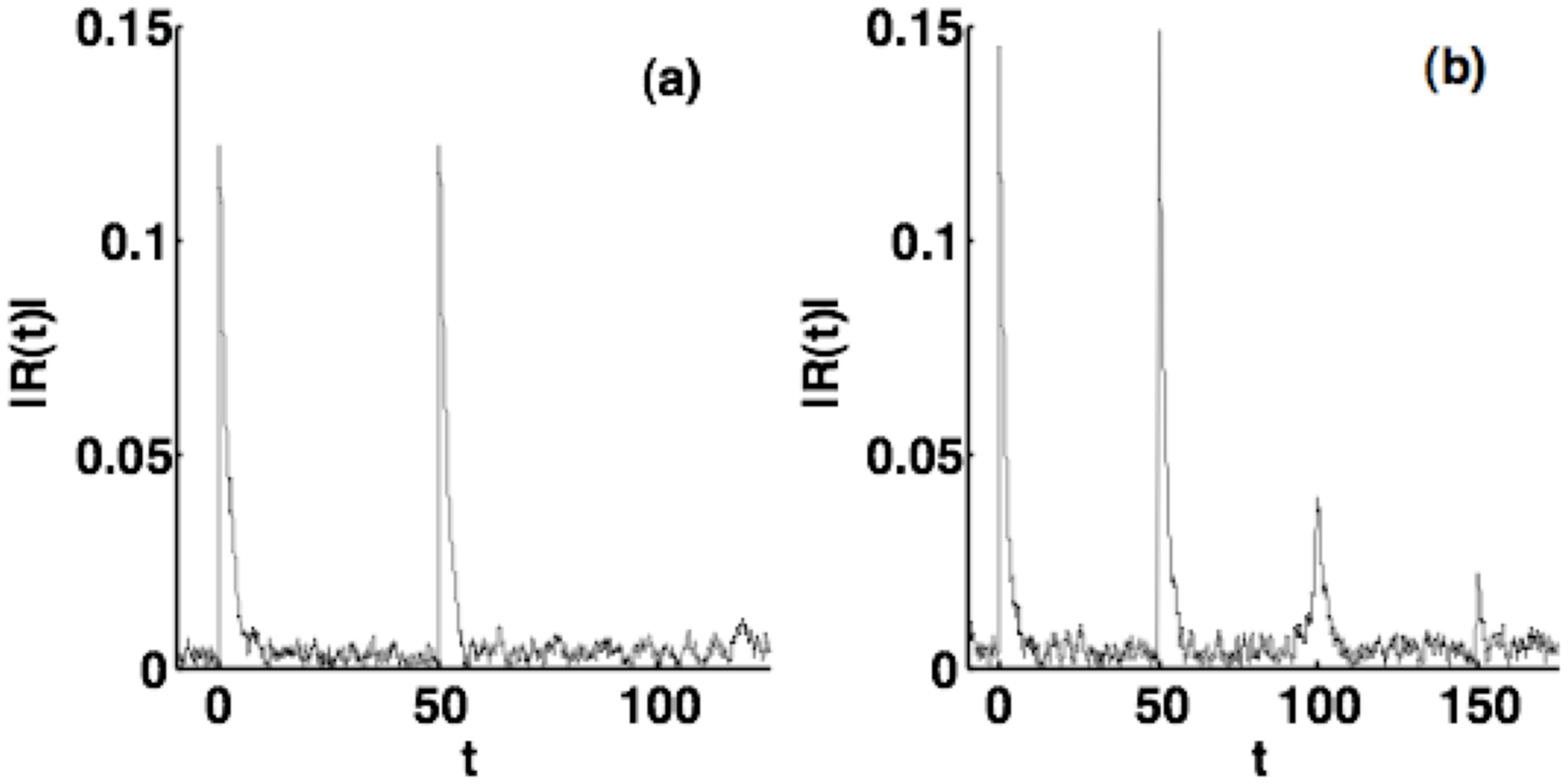}
\caption{Simulation of $10^5$ oscillators for $\tau =100$, $\hat
d_0=\hat d_1=??$, $K=1=\frac{1}{2}K_c$, $h(\theta )=\sin \theta
+\sin 2\theta +\sin 3\theta $. In this case, echoes are seen at
$t=2\tau =200$ and at $t=3\tau =300$. The inset shows a blow-up of
the numerical result for the echo shape at $t=3\tau $ with the
theoretical result, Eq.~(41), superposed (dotted curve).}
\end{figure}
Finally, we note that similar numerical experiments to all of the
above have been repeated using a Gaussian $g(\omega )$, and these
yield similar results (not shown).
\section{V. Discussion}
Echo phenomena as used for MRI provide a powerful medical
diagnostic tool. Echoes in plasmas have also been used as a basis
for measuring velocity space diffusion of plasma
particles\cite{jensen69}. Thus it is of interest to consider
whether there are potential diagnostic measurement uses of echoes
in the context of situations that can be described by the Kuramoto
model and its variants. For example, we note that the amplitude of
the echo varies exponentially with $\xi $, providing a possible
means of determining the phase diffusion coefficient $\xi $. For
example, the amplitude of the echo at $t=2\tau $ varies as
$e^{-\xi \tau }$. Thus the log of the ratio of measurements of the
echo amplitude using two different values of $\tau $, divided by
the difference in the $\tau $ values, provides a potential means
of estimating $\xi $. Also, as indicated by Eq.~(43), if one can
lower the coupling $K$ sufficiently, then echoes provide a
potential way of determining the oscillator frequency distribution
function $g(\omega )$. In particular, for low $K$ the distribution
$g(\omega )$ is directly given by the inverse Fourier transform of
the echo profile. On the other hand, we have seen from the
simulations in Sec.~IV that finite $N$ leads to noise-like
behavior that may compromise such attempts. We also note that the
Kuramoto model is an idealization, and application to any given
situation may require modifications of the model and theory to
more closely correspond to the situation at hand. We,
nevertheless, feel that consideration of echoes for diagnostics
may be of potential use.

Furthermore, these phenomena are of theoretical interest from at
least two points of view. First, as mentioned in Sec.~IIIb, the
memory required by the echo phenomenon can be thought of as
leading to a macroscopically observable consequence of the
continuous spectrum\cite{strogatz92} of the Kuramoto model. A
second point of theoretical interest relates to the recent work in
Ref.\  \cite{ott}. In that paper it was shown for a general class
of initial conditions that are on a certain manifold of the
infinite dimensional state space of the Kuramoto system, that the
future time evolution of the order parameter is determined by the
current value of the order parameter. In particular, there is an
ordinary differential equation describing the order parameter
evolution. The echo phenomenon provides an example showing that,
if initial conditions do not lie on the specified manifold of
Ref.\  \cite{ott}, other behavior can occur. In particular, well
after the second stimulus (at $t=\tau $) and well before the
occurrence of the first echo (at $t=2\tau $), the order parameter
is essentially zero, yet it does not remain zero as would be
predicted for initial conditions on the manifold of Ref.\cite{ott}
for $K<K_c$. This is discussed further in Appendix III.

In conclusion, we hope that our work will stimulate experimental
groups to investigate the type of situations we have addressed.

This work was supported by ONR (N00014-07-1-0734) and NSF
(PHY0456249).

\section*{Appendix I: Linear Analysis}

In this Appendix we solve Eq.~(15) for $0<t<\tau $ to obtain the
solution (16) and (17) for $K<K_c$. Taking the Laplace transform,
$\hat u(s)=\int ^\infty _0u(t)e^{-st}dt$, Eq.~(15) yields
\begin{equation}
\hat f_1^{(1)}(\omega ,s) =\left( \frac{K}{2}\hat
R_*^{(1)}(s)+ih_1d_0\right) /(s+\xi +i\omega )
\end{equation}
where $\hat R_*^{(1)}(s)$ denotes the Laplace transform of
$R^{(1)*}(t)$. Multiplying Eq.~(45) by $g(\omega )d\omega $ and
integrating from $\omega =-\infty$ to $\omega =+\infty$, we obtain
\begin{equation}
\hat R_*^{(1)}(s)=ih_1d_0I(s)/D(s) \ ,
\end{equation}
where $I(s)=\int ^{+\infty}_{-\infty}d\omega g(\omega )/(s+\xi
+i\omega )$ and $D(s)=1-(K/2)I(s)$. Inserting Eq.~(46) in (45)
gives
\begin{equation}
\hat f_1^{(1)}(\omega ,s)=ih_1d_0[D(s)(s+\xi +i\omega )]^{-1} \ .
\end{equation}
As noted in Sec.~IIIb, $\hat f_1^{(1)}(\omega ,s)$ has poles in
$s$ at the zeros of $D(s)$ and at $s=-(i\omega +\xi )$. These
yield time dependences of the inverse Laplace transform of $\hat
f_1^{(1)}$ (see Eq.~(27)) that vary as $e^{s_{0}t}$ and as
$e^{-(i\omega +\xi )t}$, respectively, where $s_0$ denotes the
root of $D(s)=0$ with the least negative real part. For $t\approx
\tau $ and $-[Re(s_0)+\xi ]\tau \gg 1$, we can neglect the
contributions from poles arising from roots of $D(s)=0$, and use
only the contribution from the pole at $s=-(i\omega +\xi )$. From
Eqs.~(27) and (47) this yields
\begin{equation}
f_1^{(1)}(\omega ,t)\cong ih_1d_0e^{-(i\omega +\xi )t}/D[-(i\omega
+\xi )] \ ,
\end{equation}
thus confirming Eqs.~(16) and (17).

\section*{Appendix II: Echo at $t=2\tau $ for Gaussian $g(\omega
)$} We consider the case $g(\omega )=g_G(\omega )\equiv (2\pi
\Delta ^2)^{-1/2}\exp [-(\omega -\bar \omega )^2/(2\Delta ^2)]$.
Putting this expression for $g(\omega )$ and the $n=-1$
contribution to $f_{1,e}^{(2)}$ (i.e., the first term in Eq.~(33))
into Eq.~(32) we have,
\begin{equation}
R^{(2)*}_{2\tau }(t)=\frac{h_1h_2d_0d_1}{\sqrt{2\pi \Delta
^2}}\int ^{+\infty}_{-\infty}d\omega \frac{\exp-\left\{
\frac{[(\omega -\bar \omega )+i\Delta ^2(t-2\tau )]^2}{2\Delta
^2}+\frac{\Delta ^2}{2}(t-2\tau )^2+i\bar \omega (t-2\tau )-\xi
t\right\}}{D[-(i\omega +\xi )]D^*[-(i\omega +\xi )]} \ .
\end{equation}
The collective damping rate is determined by the root of $D(s)=0$
with the least negative real part. Denote this root $s=s_0$ where
\begin{equation}
s_0=-(i\bar \omega +\xi +\gamma _0)=-(i\omega _0+\xi ) \ ,\
\omega _0=\bar \omega -i\gamma _0 \ ,
\end{equation}
where $\gamma _0>0$ is real. Letting $F(\omega )\equiv D[-(i\omega
+\xi )]$, continuing this function from real $\omega $ into the
complex $\omega $-plane, and expanding around $\omega =\omega _0$,
we have
\begin{equation}
F(\omega )=(\omega -\omega _0)\eta +\mathcal{O}[(\omega -\omega
_0)^2] \ ,
\end{equation}
where $\eta $ is a complex constant. Letting $F_*(\omega )$ denote
the continuation of the function of the real variable $\omega $ in
Eq.~(49), $D^*[-(i\omega +\xi )]$, into the complex $\omega
$-plane, we have that this function has a zero at $\omega =\omega
^*_0$,
\begin{equation}
F_*(\omega )=(\omega -\omega ^*_0)\eta ^*+\mathcal{O}[(\omega
-\omega ^*_0)^2] \ .
\end{equation}
Considering the oscillatory $\omega $ variation in the numerator
of the integrand of Eq.~(49) to be rapid (valid for $K\ll \gamma
_0$), we can approximate the integral by the saddle point method,
where the saddle point is at
\[
\omega _{sp}=\bar \omega -i\Delta ^2(t-2\tau ) \ ,
\]
and the steepest descent path through $\omega =\omega _{sp}$ runs
along the horizontal line $Im(\omega )=-\Delta ^2(t-2\tau )$ from
$Re(\omega )=-\infty$ to $Re(\omega )=+\infty$ (see Fig.~6). From
Fig.~6(a) we see that for $\Delta ^2|t-2\tau |<\gamma _0$, the
poles at $\omega =\bar \omega \pm i\gamma _0$ are not intercepted
by the steepest descent path, while for $\Delta ^2|t-2\tau
|>\gamma _0$ one of the poles is intercepted (e.g., Fig.~6(b)). In
the case where a pole is intercepted, its contribution dominates
the contribution from the saddle point by virtue of its time
dependence, $e^{-\gamma _{0}|t-2\tau |}$, as opposed to the saddle
point contribution time dependence, $e^{-\frac{1}{2}\Delta
^2(t-2\tau )^2}$. Thus we obtain
\begin{equation}
R^{(2)*}_{2\tau }(t)\sim e^{-i\bar \omega (t-2\tau )-\xi t}\times
\left\{ \begin{array}{ll} e^{-\frac{\Delta ^2}{2}(t-2\tau )^2} &
{\rm for} \ |t-2\tau |<2\gamma _0/\Delta ^2 \ , \\
e^{-\gamma _0|t-2\tau |} & {\rm for} \ |t-2\tau | >2\gamma
_0/\Delta ^2 \ . \end{array} \right.
\end{equation}
Near $\gamma _0=\Delta ^2|t-2\tau |/2$, the pole is near the
saddle point, and a uniform asymptotic expansion of the integral
(49) is necessary to obtain the transition between the two forms
in Eq.~(53).
\begin{figure}[h]
\includegraphics[scale=.75]{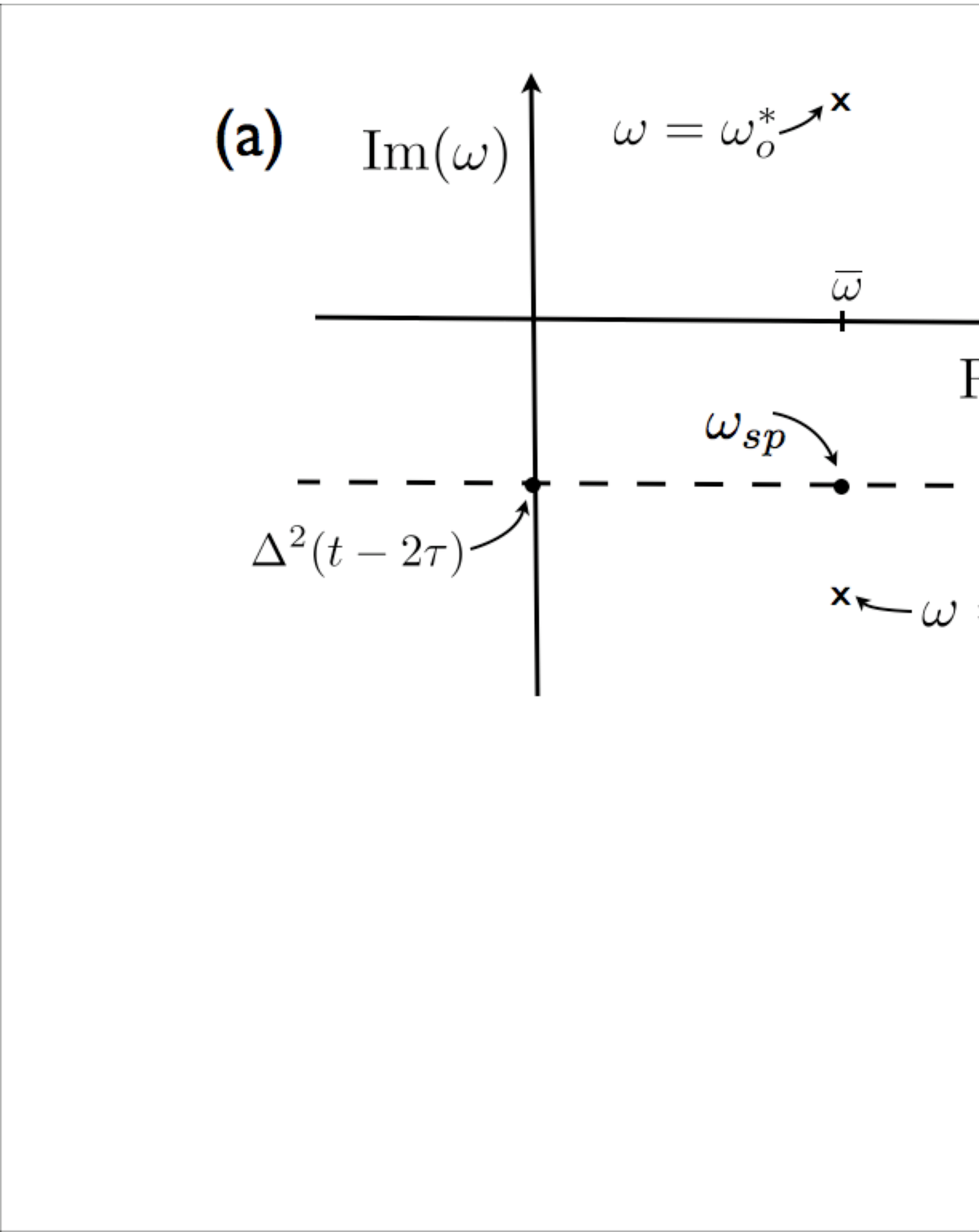}
\caption{(a) Steepest descent path (dashed) through the saddle
point $\omega =\omega _{sp}$ for $\Delta ^2|t-2\tau |<\gamma _*$.
(b) The steepest descent path (dashed) for $\Delta ^2(t-2\tau
)<-\gamma _*$. The dominant poles at the roots of $F(\omega )=0$
and $F_*(\omega )=0$ are shown as crosses, where in (b) the
steepest descent path has intercepted the pole $\omega =\omega _0$
resulting in a pole contribution to $R^{(2)*}_{2\tau }(t)$.}
\end{figure}
\begin{figure}[h]
\includegraphics[scale=.75]{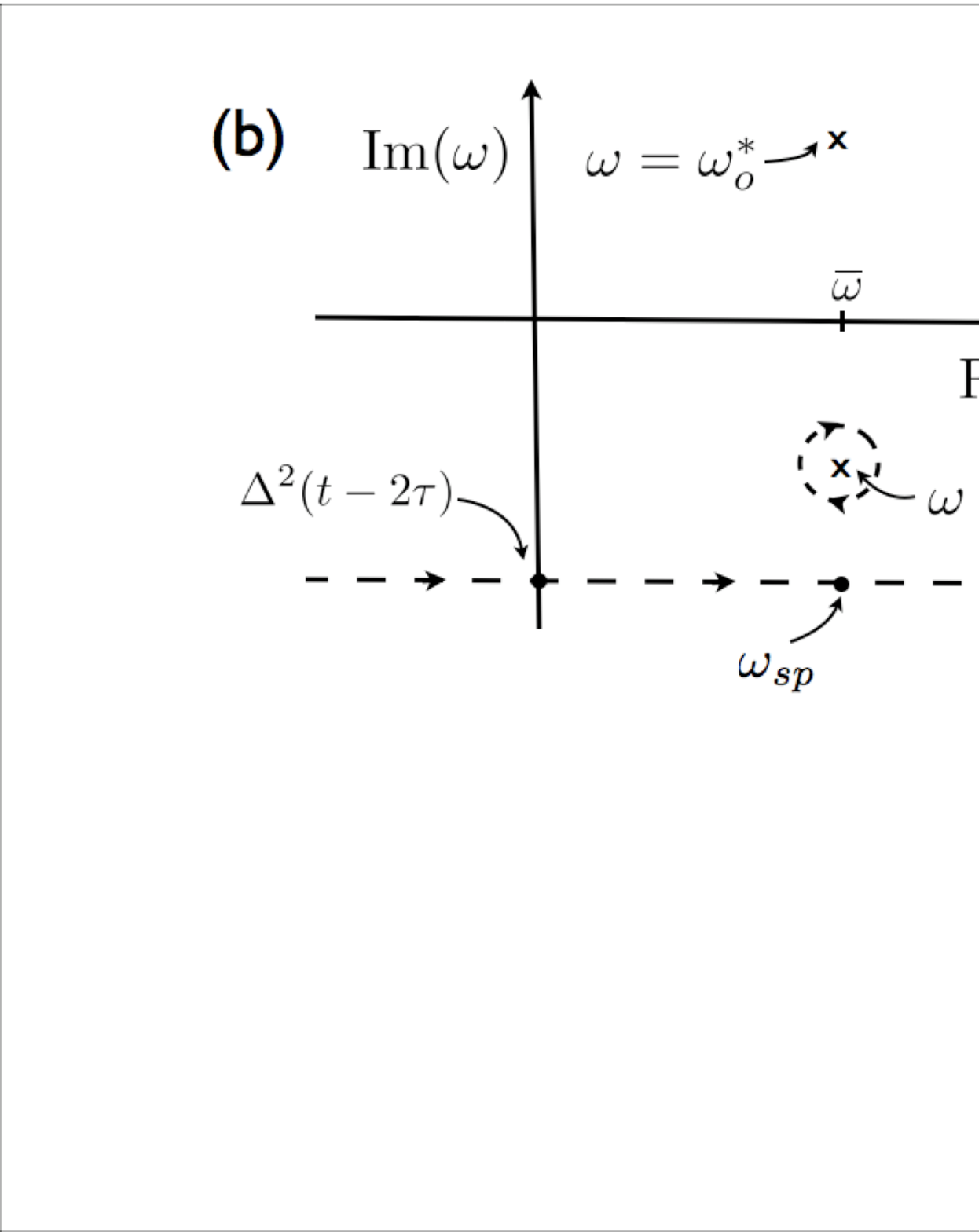}
\caption{(a) Steepest descent path (dashed) through the saddle point
$\omega =\omega _{sp}$ for $\Delta ^2|t-2\tau |<\gamma _0$. (b)
The steepest descent path (dashed) for $\Delta ^2(t-2\tau
)<-\gamma _0$. The dominant poles at the roots of $F(\omega )=0$
and $F_*(\omega )=0$ are shown as crosses, where in (b) the
steepest descent path has intercepted the pole $\omega =\omega _0$
resulting in a pole contribution to $R^{(2)*}_{2\tau }(t)$.}
\end{figure}

\section*{Appendix III:\ \ Further Discussion of Ref.\cite{ott}}

In Ref.\  \cite{ott}, a broad class of noiseless (e.g., $\xi =0$
in Eqs.~(3) and (11)) globally coupled systems of phase
oscillators was studied. The simplest example of this class is the
Kuramoto model. Reference \cite{ott} considered Lorentzian
$g(\omega )$ and a special class of initial conditions. Referring
to Eq.~(12), these initial conditions are of the form,
\begin{equation}
f_n(\omega ,0)= \alpha ^n(\omega ) \ , \
\ {\rm for } \ \ n\geq 0 \ , \\
\end{equation}
and $f_n(\omega ,0)=f^*_{-n}(\omega ,0)\ , \ \ {\rm for} \ \ n\leq
0 $\ , where $|\alpha (\omega )|<1$ for $\omega $ on the real
axis, $\alpha (\omega )$ is analytic in ${\rm Im} (\omega )< 0$,
and $|\alpha (\omega )|\rightarrow 0$ as ${\rm Im} (\omega
)\rightarrow -\infty $. Under these conditions, Ref.\ \cite{ott}
shows that the order parameters (or parameter), see Eq.~(12), that
describe the nonlinear, macroscopic time evolution of the given
system satisfy a finite set of ordinary differential equations in
time. Thus the order parameter dynamics is low dimensional, while
the dynamics of the full system determining the evolution of the
distribution function $f(\omega ,\theta ,t)$ is infinite
dimensional\cite{ott}. For example, for the Kuramoto problem with
the above conditions satisfied, Ref.\  \cite{ott} shows that
\begin{equation}
dR/dt+\left( \Delta -\frac{1}{2}K\right) R + \frac{1}{2}K\Delta
|R|^2R=0 \ ,
\end{equation}
where we have taken $\bar \omega =0$ in Eq.~(35).

A consequence of Eq.~(55) is that for $K<2\Delta \equiv K_c$,
$|R(t)|$ {\it decreases monotonically} to zero. This behavior is
not followed in the echo phenomena we discuss in the present
paper. In particular, in Fig.~1, $|R(t)|$ is small between $t=\tau
$ and $t=2\tau $, but then {\it increased} to form the echo in the
vicinity of time $t=2\tau $. Referring to Eq.~(34) and our
subsequent discussion, we see that this is because there is a
component of $f_1(\omega ,t)$ that varies as $\exp [-i\omega
(t-2\tau )]$. Identifying $f_1(\omega ,0)$ in the linear problem
with $\alpha (\omega )$ in the nonlinear problem [Eq.~(54)] and
considering $t_0$ as a new initial time (shift time so that $t_0$
goes to $t=0$), we see that $\alpha (\omega )\sim \exp [-i\omega
(t_0-2\tau )]$. If we take $t_0$ to be such that $\tau <t_0<2\tau
$ and $|R(t_0)|$ is small, then $\alpha (\omega )$ does not
satisfy the condition of Ref. \ \cite{ott} that $\alpha (\omega
)\rightarrow 0$ as $Im(\omega )\rightarrow -\infty$. However, if
$t_0>2\tau $, then it does. Thus the increase of $|R(t)|$ occurs
only when the hypothesis under which Eq.~(55) was derived does not
hold.

More generally, consider an initial condition for the original
Kuramoto problem (without stimuli or noise) where $f_1(\omega ,0)$
is analytic on the real $\omega $-axis. Expressing $f_1(\omega
,0)$ as a Fourier integral transform, we have
\begin{equation}
f_1(\omega ,0)=\int ^{+\infty}_{-\infty} e^{i\omega \eta }k(\eta )
d\eta \ ,
\end{equation}
where $k(\eta )$ is the Fourier transform of $f_1(\omega ,0)$.
Since $f_1(\omega ,0)$ is analytic in $\omega $, $k(\eta )$
decreases exponentially for sufficiently large $\eta $,
\begin{equation}
|k(\eta )|<He^{-\beta \eta } \ , \ \ {\rm if}\ \ \eta >\eta _0 \ ,
\end{equation}
for some set of positive constants $H,\beta ,\eta _0$. Using the
Laplace transform technique (as in Appendix I), it can be shown
that the solution to the linearized initial value Kuramoto problem
contains a component of $f_1(\omega ,t)$ of the form $\exp
(-i\omega t)f_1(\omega ,0)$, which we can express using Eq.~(56)
as
\begin{equation}
\exp(-i\omega t)f_1(\omega ,0)=\int ^t_{-\infty}e^{-i\omega
(t-\eta )}k(\eta )d\eta +\int ^\infty _te^{i\omega (\eta
-t)}k(\eta )d\eta \ .
\end{equation}
Setting $t=t_0$ and regarding $t=t_0$ as a new initial condition
time, we note that the initial condition consists of two terms,
namely the first and second integrals on the right hand side of
Eq.~(58). For $t_0>\eta _0$ sufficiently large, the second
integral is smaller than the first by a factor of order $\exp
(-\beta t_0)$. Furthermore, the first integral satisfies the
condition $f_1(\omega ,t_0)\rightarrow 0$ as ${\rm Im}(\omega
)\rightarrow -\infty$ [because $(\eta -t_0)>0$ for the first
integral], while the second integral does not. Thus, if we choose
to shift what we designate as the initial time to sufficiently
large $t_0$, then aside from an exponentially small component of
order $\exp (-\beta t_0)$, the initial condition obeys the
requirement of Ref.\ \cite{ott} that $f_1(\omega ,t_0)$ goes to
zero as ${\rm Im} (\omega )\rightarrow -\infty$.

\end{document}